# Intersectoral Knowledge in AI and Urban Studies: A Framework for Transdisciplinary Research

Rashid Mushkani

Université de Montréal, Mila–Quebec AI Institute

**Abstract**

Transdisciplinary approaches are increasingly essential for addressing grand societal challenges, particularly in complex domains such as Artificial Intelligence (AI), urban planning, and social sciences. However, effectively validating and integrating knowledge across distinct epistemic and ontological perspectives poses significant difficulties. This article proposes a six-dimensional framework for assessing and strengthening transdisciplinary knowledge validity in AI and city studies, based on an extensive analysis of the most cited research (2014–2024). Specifically, the framework classifies research orientations according to ontological, epistemological, methodological, teleological, axiological, and valorization dimensions. Our findings show a predominance of perspectives aligned with critical realism (ontological), positivism (epistemological), analytical methods (methodological), consequentialism (teleological), epistemic values (axiological), and social/economic valorization. Less common stances, such as idealism, mixed methods, and cultural valorization, are also examined for their potential to enrich knowledge production. We highlight how early career researchers and transdisciplinary teams can leverage this framework to reconcile divergent disciplinary viewpoints and promote socially accountable outcomes.

1.  **Introduction**

Contemporary urban challenges—ranging from climate change and public health to socioeconomic inequality—are recognized as complex, interwoven issues that demand transdisciplinary solutions (Klein, 2017). Transdisciplinary research extends beyond disciplinary silos to include non-academic stakeholders, thereby creating synergies that can yield more holistic and impactful outcomes (Frodeman et al., 2017; Hadorn et al., 2008). Within this context, AI and urban studies present an illustrative case. On the one hand, AI-driven methods promise valuable insights from large-scale datasets, while on the other, social science and urban planning contribute nuanced perspectives on community needs and societal implications (Kitchin, 2016; Koseki et al., 2022).

However, integrating diverse perspectives encounters persistent obstacles. Traditional research metrics often fail to capture the complexity inherent in transdisciplinary endeavors, making it difficult to validate or fund such work (Bromham et al., 2016; Wagner et al., 2011). Researchers trained in computer science and engineering may emphasize computational rigor and quantitative methods, whereas those in social science and urban planning highlight contextual, qualitative insights (Hadorn et al., 2008). Without a common evaluative framework, translating knowledge across these paradigms can be fraught with miscommunication, methodological incompatibilities, and limited reproducibility (Rousseau et al., 2019).

Against this backdrop, the current study aims to:
- Identify dominant and less frequently employed perspectives in AI and urban research across six dimensions (ontological, epistemological, methodological, teleological, axiological, and valorization).
- Propose a framework that helps researchers select and align perspectives in ways that foster credibility and societal impact in transdisciplinary collaborations.
- Demonstrate how this framework can guide intersectoral research design, helping practitioners and policymakers make informed decisions about suitable epistemic and methodological pathways.

By mapping widely cited research on AI and urban studies, we advance the conversation on how to validate knowledge in contexts that integrate multiple sectors and disciplines. This study also highlights approaches to overcome barriers posed by disparate epistemologies, thus offering practical guidance for more inclusive, socially accountable research.

## 2. Background
### 2.1. Intersectorality and transdisciplinarity

Intersectoral and transdisciplinary research strives to move beyond traditional disciplinary confines by actively involving stakeholders from industry, government, non-profit organizations, and local communities (Klein, 2017). In contrast to multidisciplinary or interdisciplinary work, transdisciplinarity explicitly emphasizes incorporating non-academic actors in knowledge creation and problem-solving (Frodeman et al., 2017). This broader scope is essential for tackling "grand challenges" such as sustainable urban development, as these issues inherently span multiple sectors (Bromham et al., 2016; Hadorn et al., 2008).

Yet, the structural and cultural barriers that hinder cross-sector collaboration remain substantial. Metrics and review processes tailored to narrowly defined disciplinary outputs can undervalue transdisciplinary research (Wagner et al., 2011). Researchers also face difficulties in harmonizing different vocabularies, methods, and objectives. Consequently, efforts to reorganize academic and funding structures around intersectoral goals are gaining momentum, with increased calls for revised evaluation criteria and the expansion of specialized outlets for publishing cross-sector studies (Garner et al., 2013; Hu et al., 2024).

### 2.2. Urban complexity

Cities are multifaceted entities encompassing physical infrastructure, social systems, and cultural dynamics (Batty, 2013; Mumford, 1968). They can be studied through numerous lenses: spatial organization, social interactions, economic factors, and environmental impacts, among others (Bondi, 1998; Juhász & Hochmair, 2016). Historically, urban studies have evolved from qualitative approaches—focused on descriptive and interpretive analyses of human behavior in cities—to more sophisticated quantitative and computational models that recognize urban areas as complex adaptive systems (Batty, 2018; Portugali, 2011).

As digital technologies become ever more integrated into urban life, the potential for data-driven urban science expands. Large-scale datasets can uncover new insights into mobility patterns, land use, or environmental stressors (Kitchin, 2016; Koseki et al., 2022). However, purely quantitative methods can overlook socio-cultural dimensions that shape urban life, such as power structures and cultural identities (Low, 2020). Incorporating social-scientific understanding is thus essential for capturing the full depth of urban phenomena, ensuring that interventions and solutions are contextually grounded (Calhoun, 2017).

### 2.3. Computer science in urban studies

Computer science, including its subfields of AI and data analytics, has increasingly intersected with urban research (Batty, 2018; Kitchin, 2016). Initial conceptions of computer science were heavily rooted in logical and mathematical traditions (Dahl et al., 1972), but the field has since broadened to accommodate empirical testing, engineering approaches, and, more recently, advanced AI algorithms (Angius et al., 2024).

The application of AI to urban problems spans traffic flow analysis, predictive maintenance of infrastructure, urban logistics, and beyond (Alzubaidi et al., 2023; Koseki et al., 2022). Yet challenges arise when computational methods are applied to social contexts without acknowledging local realities or ethical considerations (Benjamin, 2019; Costanza-Chock, 2020). This underscores the importance of coupling the technical strengths of computer science with participatory, reflexive, and transdisciplinary frameworks for knowledge production (Hadorn et al., 2008).

### 2.4. Urban science and informatics

Urban science (or city science) and urban informatics illustrate the convergence of computational tools with spatial and social approaches to city studies (Batty, 2013; Kitchin, 2016). These fields often champion data-driven insights, drawing on large, heterogeneous datasets to inform policy or planning (Portugali, 2011). While such computationally intensive methods can be powerful, they may run the risk of determinism or reductionism if they neglect qualitative inputs regarding community values, political contexts, and historical legacies (Batty, 2018; Kitchin, 2016).

An emerging consensus argues that successful urban informatics must be transdisciplinary, integrating stakeholder participation from the earliest stages to design data systems and analyses that genuinely benefit diverse urban populations (Rousseau et al., 2019). In this manner, the sociotechnical complexity of urban environments can be addressed with methodological pluralism, combining advanced computation, social engagement, and ethical reflexivity (Hidalgo, 2016; Kamrowska-Załuska, 2021).

### 3. Theoretical framework

Intersectoral research emphasizes the co-construction of knowledge by involving research users and beneficiaries throughout the decision-making process. This approach aligns with ethical goals such as self-determination and cognitive justice (Hadorn et al., 2008; Pratt, 2019). Community-Based Participatory Research, for example, partners with marginalized groups to address social disparities, thereby increasing the relevance and equity of research outcomes (Fischer, 2000; Gibbons et al., 1994; Pratt, 2019).

Nevertheless, developing intersectoral research proposals can be challenging, particularly in grant evaluation contexts that favor conventional disciplinary approaches. Explaining complex, multi-field methodologies to review panels or linking novel frameworks to traditional funding criteria often proves difficult (Bromham et al., 2016). Consequently, revised evaluation strategies and metrics tailored to intersectoral research are needed (Hadorn et al., 2008; Klein, 2008).

Achieving effective intersectoral collaboration requires significant "pre-problematization," the formation of a shared language, and the transfer of concepts across fields (Hidalgo, 2016; Klein, 2017). While the development phase may be prolonged by differences in vocabularies, analytic methods, and outcome interpretations, such efforts typically yield more innovative and comprehensive results (Vantard et al., 2023). Early engagement in intersectoral projects can foster deeper, long-term commitment among researchers, underscoring the importance of involving participants from the outset (Vantard et al., 2023).

Measuring knowledge transfer across disciplines entails assessing dimensions such as broadness, intensity, and homogeneity (Rousseau et al., 2019). Broadness refers to how frequently a given discipline's outputs are cited across other fields, intensity captures the proportion of cross-disciplinary citations, and homogeneity gauges the cognitive similarity between the citing and cited disciplines (Zhou et al., 2023). Such metrics can clarify the depth and impact of intersectoral collaborations.

Based on these considerations, the current study proposes a framework to validate how knowledge emerging from AI- and urban-focused research can be made acceptable to multiple disciplinary and societal stakeholders. By analyzing the most cited literature from the past decade in AI and urbanism, the framework identifies perspectives that are widely adopted as well as those less validated, thereby guiding researchers in selecting appropriate methodological and epistemological orientations to address complex urban challenges effectively.

Comprehensively engaging urban studies, computer science, and AI requires attention to multiple dimensions. Ontological perspectives clarify the nature of the phenomena being investigated (Gibbons et al., 1994; Veber, 2014). Realism posits that external objects and phenomena exist independently of human perception; Idealism holds that reality is mentally constructed (Chakravartty, 2017). Materialism focuses on physical substances, while Dualism separates the mental from the physical (Psillos, 2022; Robinson, 1982). Critical realism, developed by Bhaskar (1975), recognizes an objective reality yet acknowledges that social and historical contexts mediate how it is understood (Chakravartty, 2017). Bounded relativism suggests that interpretations are context-bound, but shaped by observable structures (Hammersley, 1990; Psillos, 2022).

Epistemological stances inform how knowledge is validated. Empiricism emphasizes observation and experiment (Henderson, 2022); Rationalism posits that reason can reveal truths that observation alone may not capture (Chakravartty, 2017). Constructivism asserts that knowledge is shaped through interaction with cultural or historical contexts (Liu & Matthews, 2005). Positivism views scientific, mathematically treated knowledge as the only legitimate form (Bourdeau, 2023; Turri et al., 2021).

Methodological choices stem from these ontological and epistemological commitments (Creswell, 2013). Quantitative methods rely on statistical or mathematical techniques for generalizable findings (Creswell & Creswell, 2022); Qualitative methods explore meanings and experiences through interviews or ethnographic observation (Denzin & Lincoln, 2011). Mixed methods combine quantitative and qualitative approaches for pragmatic reasons (Fortin & Gagnon, 2010). Experimental methods manipulate variables to test cause-and-effect, whereas analytical methods use logical or conceptual analysis (Beaney, 2001; Creswell & Creswell, 2022).

Teleological considerations focus on the purposes and goals of research (Ginsborg, 2022). Consequentialism evaluates actions by outcomes, while deontological perspectives emphasize duties independent of results (Alexander & Moore, 2021). Virtue ethics centers on character development (Hursthouse & Pettigrove, 2023), whereas pragmatism examines the practicality and effects of ideas (Legg & Hookway, 2021).

Axiological analysis addresses values and principles guiding research (Ginsborg, 2022; Gustafsson, 2020). Ethical values concern moral rightness, aesthetic values involve perceptions of beauty or creativity (Guyer, 2005; Hursthouse & Pettigrove, 2023), and epistemic values foreground truth and objectivity (Guyer, 2005; Schroeder, 2021).

Valorization involves recognizing and enhancing the broader value of research outputs. Economic valorization focuses on translating findings into market or industry benefits (Etzkowitz & Leydesdorff, 2000; Schroeder, 2021). Social valorization emphasizes improving societal well-being, often through policy recommendations or direct community interventions (Wolfensberger, 1983). Cultural valorization supports the preservation or promotion of cultural practices and diversity (Meissner, 2017).

These six dimensions collectively offer a structured perspective on how knowledge is generated, interpreted, applied, and valued in transdisciplinary settings. By explicating the varying ontological, epistemological, methodological, teleological, axiological, and valorization stances, the proposed framework assists researchers and stakeholders in navigating the complexities of AI- and urban-oriented research collaborations.

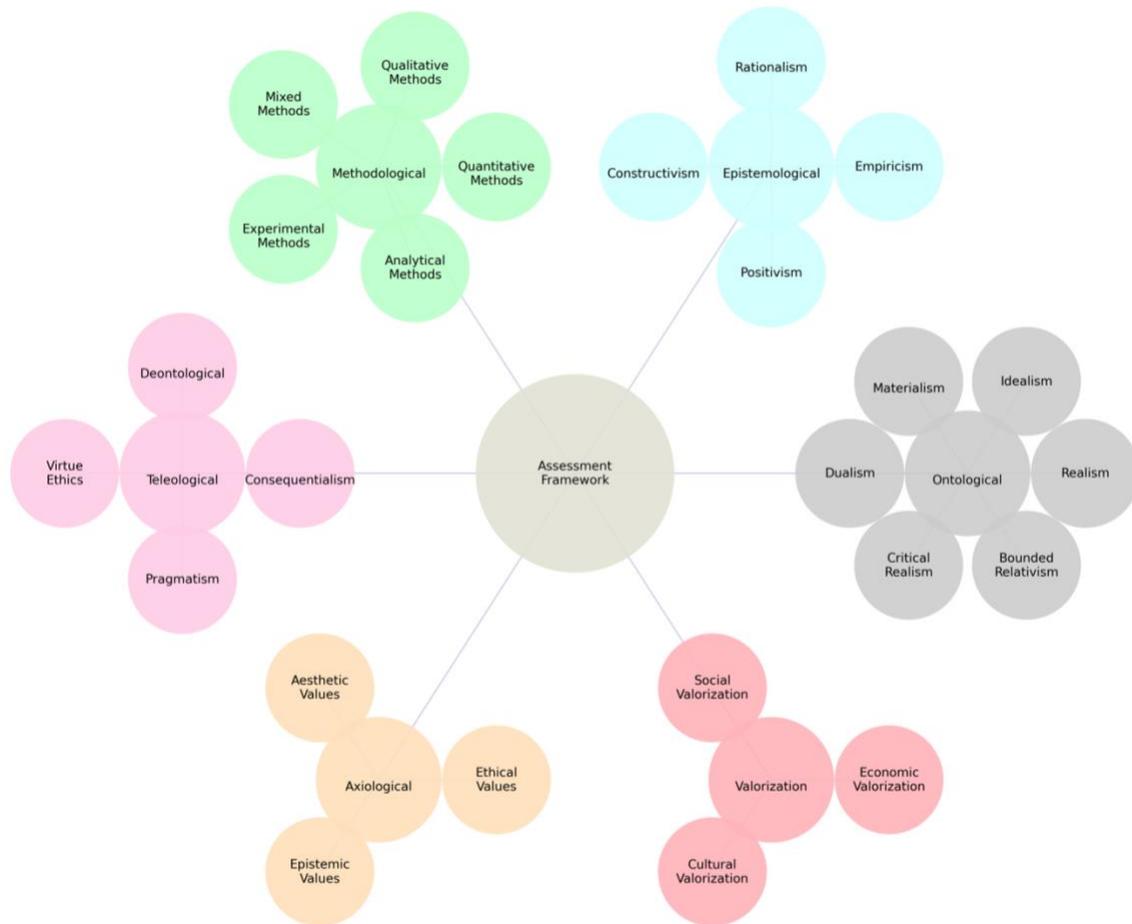

*Figure 1. Six dimensions (ontological, epistemological, methodological, teleological, axiological, and valorization) that guide transdisciplinary knowledge validation.*

### 4. Methodology and Materials

This section describes the processes undertaken to select, screen, and classify the literature. The objective was to identify influential work spanning four broad domains—AI in computer science, city in computer science, city in social science, and AI in social science—so as to capture diverse perspectives on urban and AI-related research.

#### 4.1. Data Source and Timeframe

A literature search was performed using the Scopus database for publications dated 2014–2024. The year 2014 was chosen to capture developments emerging in the mid-2010s, when AI and urban analytics began to intersect more frequently, while 2024 was included to account for recent or forthcoming work. It is acknowledged that data for 2024 may be provisional or incomplete; however, including publications indexed with early publication dates helped capture current trends. All articles were restricted to English-language records, which may exclude relevant research in other languages. Although this limitation may reduce global coverage, it aligns with commonly used international academic search strategies in these fields.

### 4.2. Search Strategy and Scope

Relevant keywords were developed to ensure coverage of each domain's thematic focus (e.g., "AI," "machine learning," "urban planning," "smart city," "AI ethics"). Pilot searches were conducted to refine terms and ensure they returned sufficiently broad yet targeted results. The final search strings were applied separately to:

1. AI in computer science
2. City in computer science
3. City in social science
4. AI in social science

These searches generated initial article sets of varying sizes (ranging from approximately 7,900 to 47,000 records).

### 4.3. Inclusion and Exclusion Criteria

The dataset encompassed conference papers, journal articles, reviews, book chapters, and books to reflect different dissemination practices across fields. Articles were limited to those published in English from 2014 onward. An initial screening excluded duplicates, ensuring that reprints or multiple records of the same work were consolidated. Subsequently, only the 500 most cited items in each domain were retained for further analysis. Citation counts were taken directly from Scopus without normalization across publication years, recognizing that older publications may accumulate more citations than recent ones. However, choosing a fixed number (500) in each domain helped maintain a consistent sample size and mitigated some disciplinary differences in citation practices.

Literature reviews were excluded on the grounds that the study aimed to analyze original research contributions rather than syntheses of existing work. This decision may omit certain conceptual or theoretical advances presented in review articles, but it was deemed appropriate to focus on primary research outputs when assessing epistemological and methodological orientations.

Table 1 Literature Collection Methodology Table

| Domain | AI in computer science | City in computer science | City in social science | AI in social science |
|---|---|---|---|---|
| Search Conditions | Limited to computer sciences: 39,960 articles | Limited to computer sciences: 12,760 articles | Limited to social sciences: 47,013 articles | Limited to social sciences: 7,904 articles |
| Keywords | "AI (artificial intelligence)", "ML (machine learning)", "DL (deep learning)", "neural networks" "reinforcement learning", "computer vision" | "city", "urban planning", "smart cities", "smart city", "urban design", "cities", "urban development", "urbanization", "public space", "urban area" | "city", "urban area", "urban planning", "smart city", "urban development", "urbanization", "urban area", "sustainable development", "public space", | "AI (artificial intelligence)", "ML (machine learning)", "DL (deep learning)", "AI ethics", "responsible AI", "inclusive AI", "fairness AI" |
| Why | Covers a wide range of AI research in computer science, focusing on high-impact works. | Captures research on urban planning and smart cities within the computer science domain, highlighting influential studies. | Provides insight into social science perspectives on urban studies, emphasizing significant contributions. | Highlights the intersection of AI and social sciences, focusing on impactful research. |
| Document types | Conference Papers, Articles, Reviews, Book Chapters, Books | | Articles, Conference Papers, Book Chapters, Reviews, Editorials | |
| Language, time frame, final screening, and exported data: English, 2014-2024, 500 most cited papers, title, abstract, keywords, year, source title, publisher, sponsor, affiliation, citation count | | | | |

The dissemination mediums varied, with conference proceedings being the predominant medium for AI in computer science and city in computer science, while journals were more prevalent in city in social science and AI in social science. This highlights the preference for rapid dissemination of findings in the computer science fields through conferences, whereas the social sciences rely more on journal publications for peer-reviewed research dissemination.

The summarized data in Tables 1 and 2 outlines the methodology for literature collection and compares the dissemination mediums across the four domains. This analysis ensures that the selected literature represents a broad spectrum of research in AI and urbanism.

Table 2. Comparison Across Four Domains

| Domain | Dissemination Mediums | Major Funders |
|---|---|---|
| AI in computer science / city in computer science | Conference proceedings, journals, book series, books, trade journals | 1. National Natural Science Foundation of China<br><br>2. National Science Foundation (US)<br><br>3. Horizon 2020 Framework Programme (EU) |
| AI in social science / city in social science | Journals, conference proceedings, book series, books, trade journals | |

### 4.4. Data Extraction

For each retained article, bibliographic details (title, abstract, keywords, year, source title, publisher, sponsor, affiliation, funding text, citation count) were exported. Funding bodies were identified to note potential influences on research agendas, and differences in dissemination outlets—journals versus conference proceedings—were examined to understand domain-specific publication practices. After applying the stated criteria, the dataset comprised 2,000 articles (4 domains × 500 articles each).

### 5. Analysis and findings

This study aimed to identify perspectives that enhance the validity of AI and city research in social science and computer science. In line with the research questions, 500 of the most cited papers were selected for each of the four key domains—AI in social science, AI in computer science, city in social science, and city in computer science—yielding 2,000 total articles. Each paper was then classified along six analytical dimensions (ontological, epistemological, methodological, teleological, axiological, and valorization), producing 12,000 dimension-specific evaluations.

A systematic approach was developed to categorize each research article. Input parameters—title, keywords, abstract, and a predefined set of potential categories—were analyzed to determine which classification best matched the content of each dimension. The GPT-4o Large Language Model with a 128k context window facilitated this task (OpenAI, 2024). For instance, the prompt structure included:

> [INST] Given the research titled '{title}' with keywords '{keywords}'

and the following abstract: '{abstract}', identify which category
among {', '.join(categories)} best describes the {dimension} dimension
of the research. Provide only the category name. If unclear or unsure,
respond with 'not classifiable'. [/INST]

Manual checks were performed where the model's classification seemed unclear, ensuring a consistent and reliable categorization (OpenAI, 2024).

The dataset was then examined to determine the most frequently occurring classification in each dimension across the four domains, enabling an understanding of dominant views in the literature (Pizzorno & Berger, 2023). A coverage analysis calculated the proportion of articles within each domain that aligned with specific classifications. This helped identify how widely certain perspectives are utilized, thereby supporting or challenging their credibility across AI- and city-focused studies.

The representative classifications identified were critical realism (ontological), positivism (epistemological), analytical methods (methodological), consequentialism (teleological), epistemic values (axiological), and social valorization (valorization). These classifications were chosen due to their consistent predominance across the sampled literature, suggesting they function as mainstream approaches.

The results of the coverage analysis showed that these prevalent perspectives enjoyed substantial representation: AI in social science (0.67), AI in computer science (0.66), city in social science (0.63), and city in computer science (0.69). Such coverage supports the argument that adopting these perspectives may enhance the perceived validity and acceptance of intersectoral research in AI and urban contexts.

Beyond these dominant perspectives, the study also identified the least common classifications, indicating positions that may be less established or carry greater methodological or epistemic risk. These less frequent stances were idealism (ontological), rationalism (epistemological), mixed methods (methodological), pragmatism (teleological), aesthetic values (axiological), and cultural valorization (valorization). Their coverage rates were comparatively low—ranging from 0.05 to 0.09—suggesting a narrower base of validation across the four domains.

Network analysis further illuminated how these classifications co-occur within articles (Hagberg et al., 2008). For each dimension, co-occurrences of classifications were extracted and visualized using a network graph, with edges representing the strength (frequency) of their association. Degree centrality identified classifications that strongly connect with others, pointing to conceptual clusters within the literature (Perrone et al., 2020). Larger nodes—such as "empiricism," "pragmatism," and "critical realism"—signified higher centrality, implying that these perspectives are frequently combined across domains and dimensions. Edges with greater thickness indicated robust co-occurrence relationships, emphasizing that certain methodological and epistemological stances appear together more often, which might boost their collective credibility in interdisciplinary work on AI and urbanism. The color-coding of nodes by dimension served to highlight the variety and distribution of perspectives.

Overall, these findings suggest that while critical realism, positivism, analytical methods, consequentialism, epistemic values, and social valorization dominate the discourse, there remains scope for integrating less common perspectives that may bring additional depth or address context-specific needs in AI and city research.

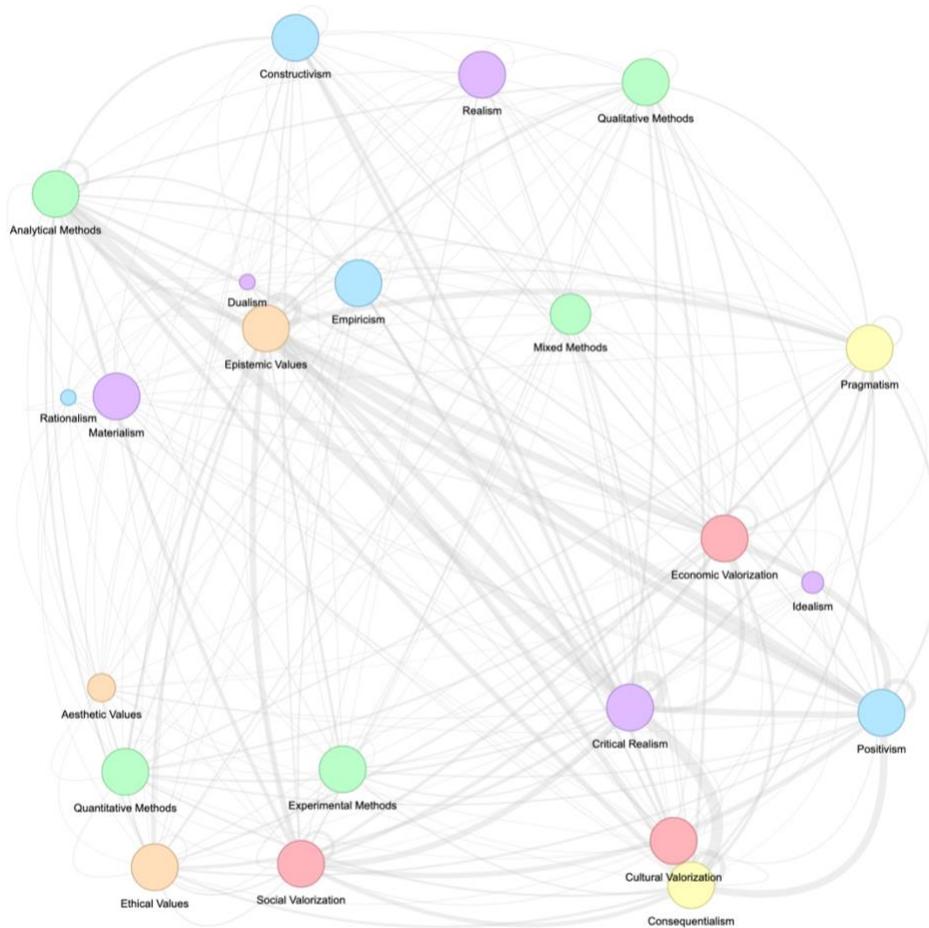

Figure 1. Network visualization with nodes representing classifications and edges representing the strength of co-occurrences

### 5.1. Ontological dimension

The analysis of ontological perspectives revealed that critical realism is the prevailing perspective across all domains. In AI in computer science, critical realism dominates with 424 instances, followed by materialism (35), realism (32), dualism (6), and idealism (3). Bounded relativism is notably absent. AI in social science emphasizes critical realism (480), with realism (4), idealism (2), bounded relativism (3), and materialism (11) being less prominent. Dualism is absent. In city in computer science, critical realism is dominant with 434 instances, followed by realism (30), materialism (34), and idealism (2). Bounded relativism and dualism are absent. City in social science shows critical realism (470) as the prevailing perspective, with materialism (18), realism (10), idealism (2), dualism and materialism being less emphasized. These insights suggest that critical realism provides a robust framework for understanding the structures and mechanisms underlying observed phenomena in AI and urban studies, making it a valuable perspective for future research.

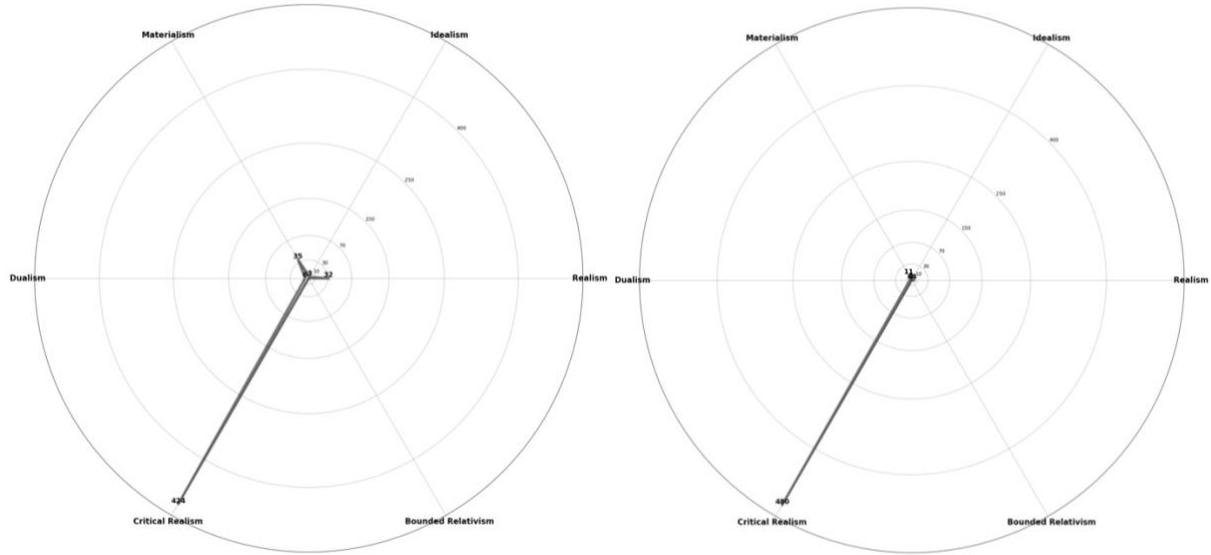

Figure 2. Distribution of ontological perspectives. Left: AI in computer science. Right: AI in social science.

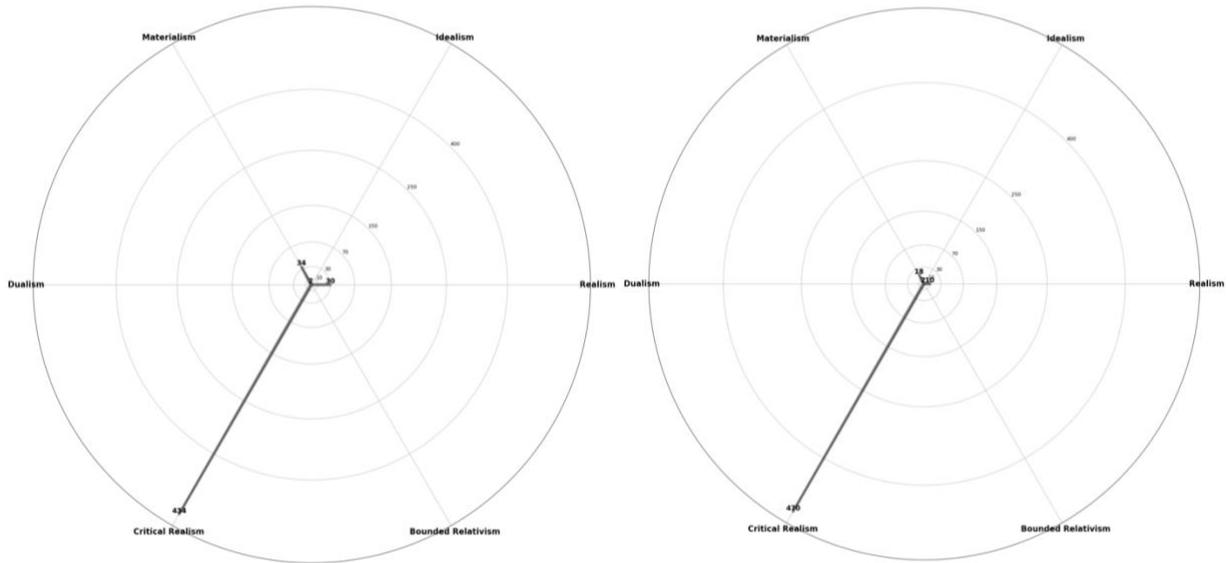

Figure 5. Distribution of ontological perspectives. Left: City in computer science. Right: City in social science.

### 5.2. Epistemological dimension

The analysis indicates that positivism is the dominant epistemological perspective across all domains. In AI in computer science, positivism dominates with 288 instances, followed by empiricism (125), constructivism (84), and rationalism (3). AI in social science prioritizes positivism (275) but shows a significant presence of constructivism (168), indicating a more balanced approach with moderate empiricism (54) and minimal rationalism (3). In city in computer science, positivism is overwhelmingly dominant with 409 instances, with less emphasis on empiricism (68) and constructivism (23), and no presence of rationalism. City in social science shows a blend of positivism (275) and constructivism (155), with empiricism (67) playing a role and rationalism remaining minimal (3). These findings highlight the dominance of positivism, especially in computer science contexts, and the balanced presence of constructivism in social sciences. This suggests the need to integrate diverse epistemological approaches in AI and urban studies to ensure research is both empirically rigorous and socially relevant.

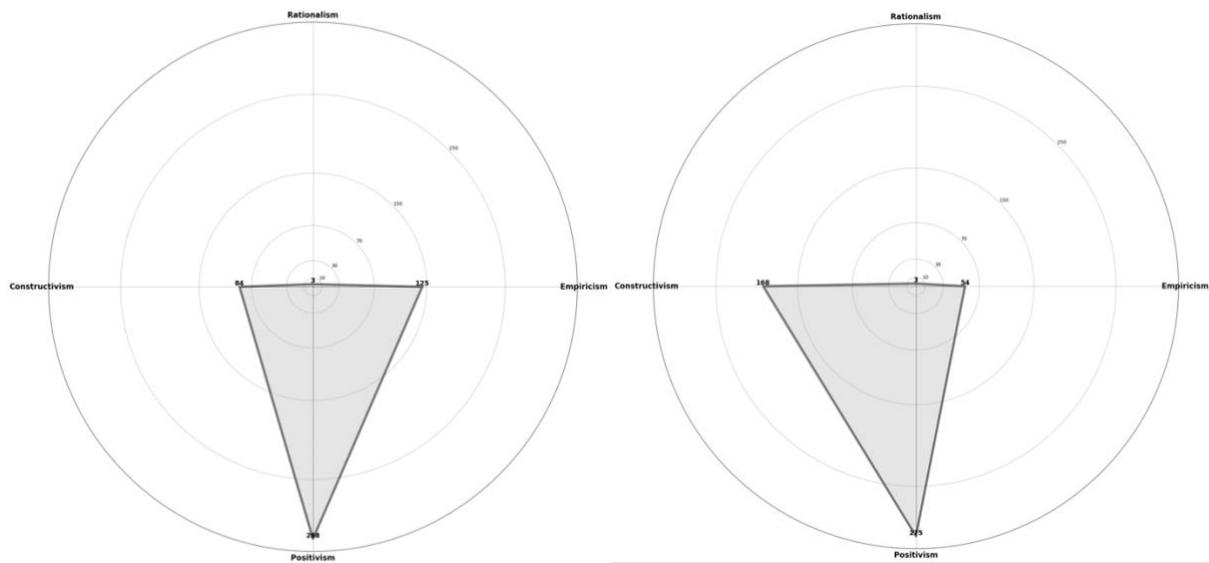

*Figure 3. Distribution of epistemological perspectives. Left: AI in computer science. Right: AI in social science.*

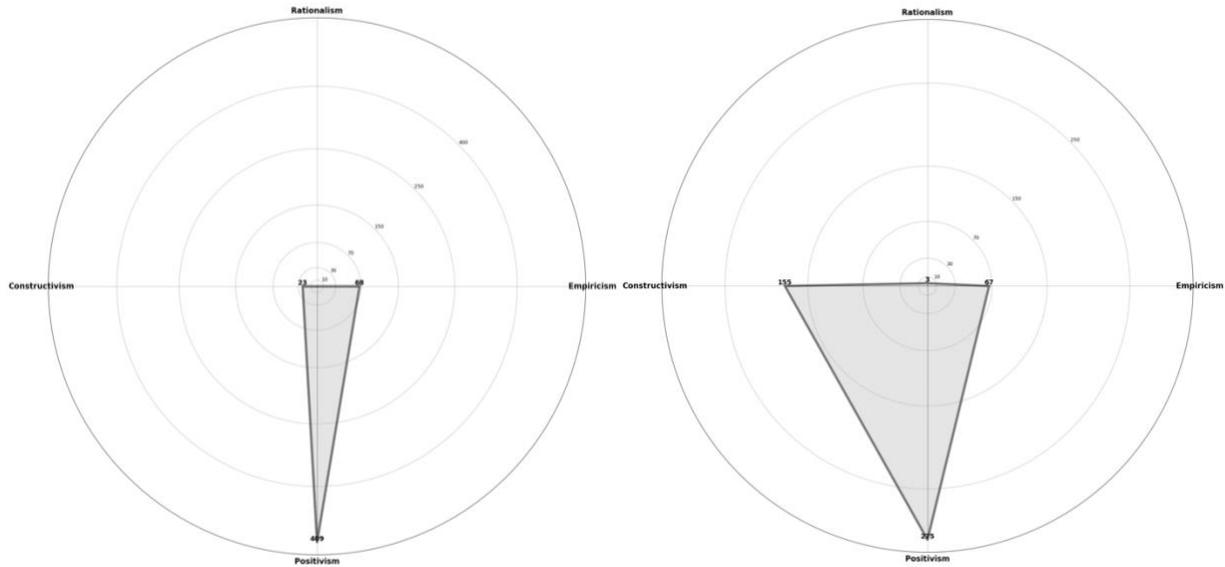

*Figure 4. Distribution of epistemological perspectives. Left: City in computer science. Right: City in social science.*

### 5.3. Methodological dimension

The analysis shows that analytical methods are the dominant methodological perspective across AI and city studies in computer science, while social science domains demonstrate a greater acceptance of qualitative methods. In AI in computer science, analytical methods dominate with 315 instances, followed by quantitative methods (63), experimental methods (69), and qualitative methods (33). Mixed methods are the least represented with 20 instances. AI in social science shows a more varied distribution, with analytical methods (227) and qualitative methods (127) being prominent. Quantitative methods (64), experimental methods (42), and mixed methods (40) are also present. In city in computer science, analytical methods again dominate with 324 instances, followed by quantitative methods (90), experimental methods (52), and qualitative methods (25). Mixed methods are minimally represented with 9 instances. City in social science shows a more evenly distributed methodological perspective, with qualitative methods (166) and analytical methods (182) being the most prominent. Quantitative methods (107), experimental methods (11), and mixed methods (30) are also represented. These findings highlight the dominance of analytical methods in computer science and the balanced use of qualitative methods in social sciences. This suggests the need to integrate various methodological perspectives in AI and urban studies to address research questions effectively.

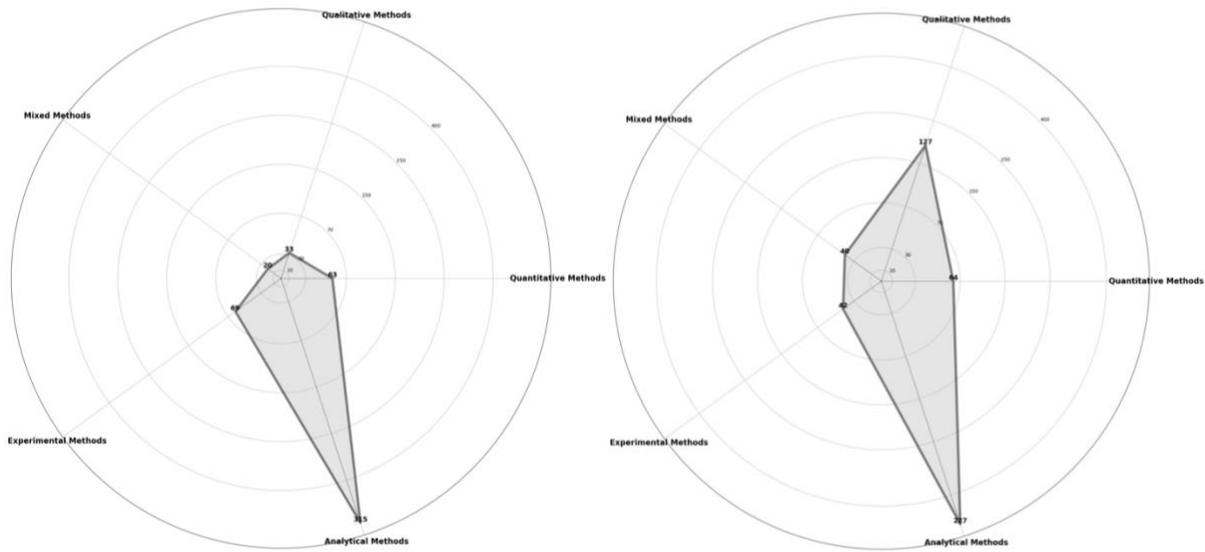

*Figure 5. Distribution of methodological perspectives. Left: AI in computer science. Right: AI in social science.*

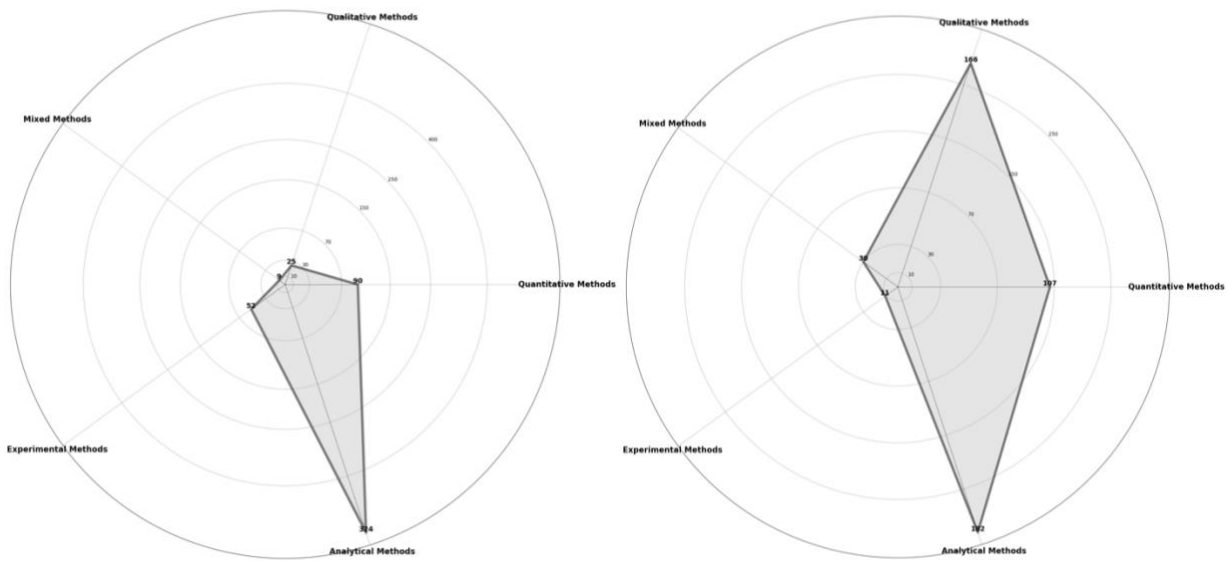

*Figure 6. Distribution of methodological perspectives. Left: City in computer science. Right: City in social science.*

### 5.4. Teleological dimension

Consequentialism is the dominant teleological perspective across all domains, with pragmatism also significant, especially in urban studies. In AI in computer science, consequentialism is the dominant perspective with 325 instances, followed by pragmatism (175). Virtue ethics and deontological perspectives are absent. AI in social science prioritizes consequentialism (410), but pragmatism is less prominent with 90 instances, and virtue ethics is absent. In city in computer science, consequentialism again dominates with 268 instances, followed by pragmatism (212). Virtue ethics and deontological perspectives are absent. City in social science shows a slightly different pattern with consequentialism (312) and pragmatism (188) being the main perspectives. These insights indicate that focusing on the outcomes of actions and adopting pragmatic approaches are essential for addressing complex issues in AI and urban studies, ensuring that research has practical relevance and impact.

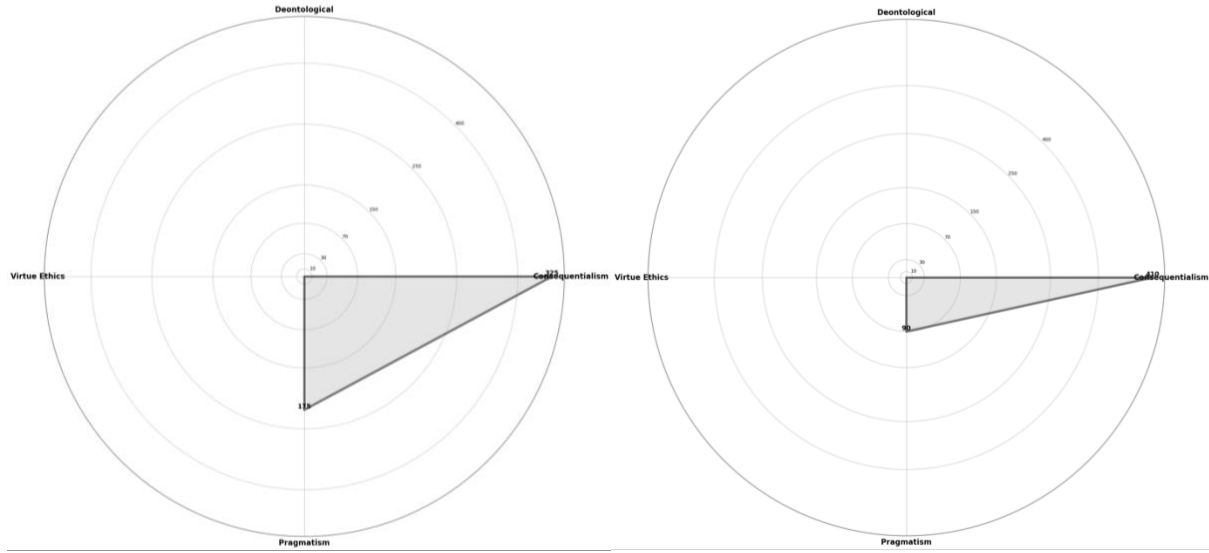

Figure 7. Distribution of teleological perspectives. Left: AI in computer science. Right: AI in social science.

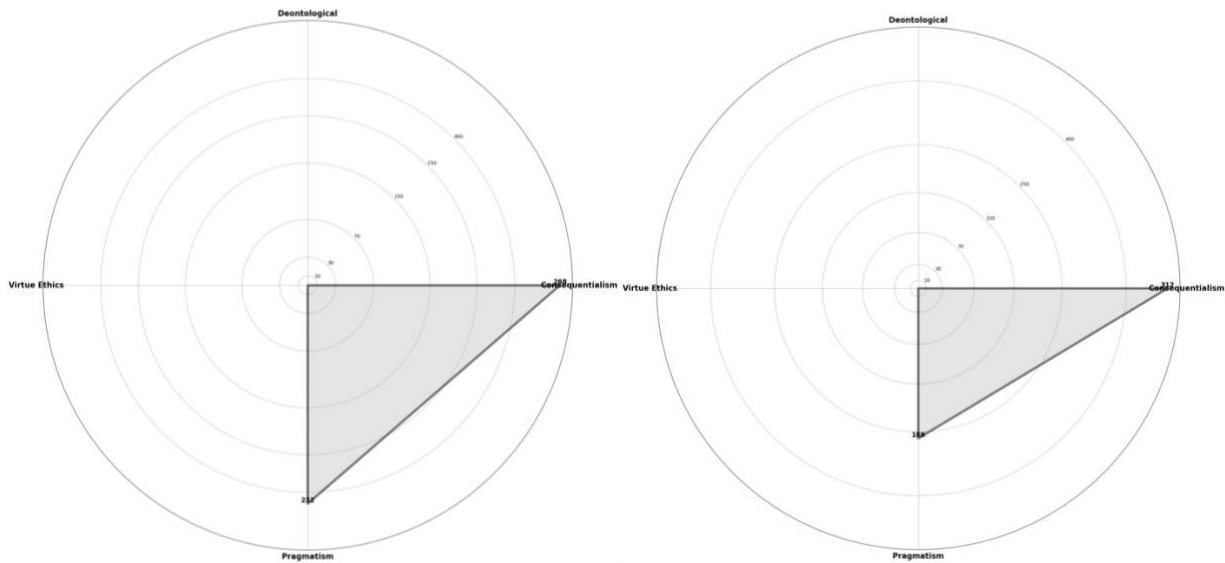

Figure 8. Distribution of teleological perspectives. Left: City in computer science. Right: City in social science.

### 5.5. Axiological dimension

The analysis of axiological dimensions—epistemic, ethical, and aesthetic values—shows that epistemic values dominate across all domains, with variation in ethical values and minimal presence of aesthetic values. In AI in computer science, there is a strong emphasis on epistemic values (410 instances), with ethical values (84) also playing a significant role, while aesthetic values are minimal. AI in social science shows a more balanced approach between epistemic (312) and ethical values (185), with aesthetic values (3) being least prioritized. City in computer science prioritizes epistemic values (459), with lower ethical values (38) and minimal aesthetic values (3). City in social science presents a nuanced distribution with epistemic values (390) dominating, followed by ethical values (97) and slightly higher aesthetic values (13) compared to other domains. These insights highlight the importance of prioritizing epistemic values to ensure the credibility and truthfulness of research, while also considering ethical implications to address societal impacts.

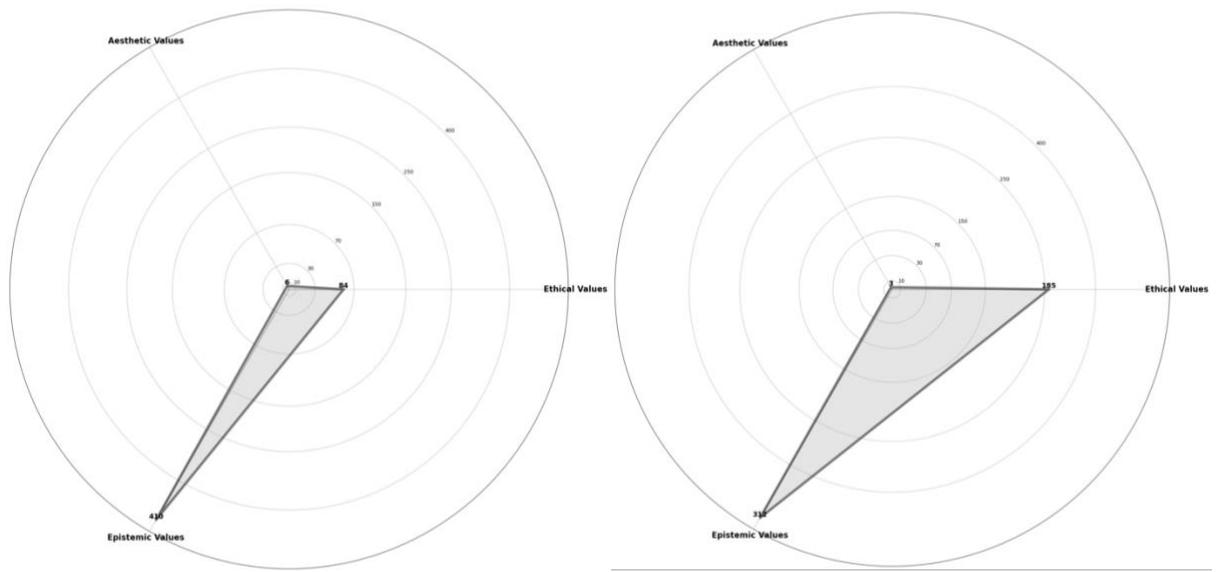

*Figure 9. Distribution of axiological perspectives. Left: AI in computer science. Right: AI in social science.*

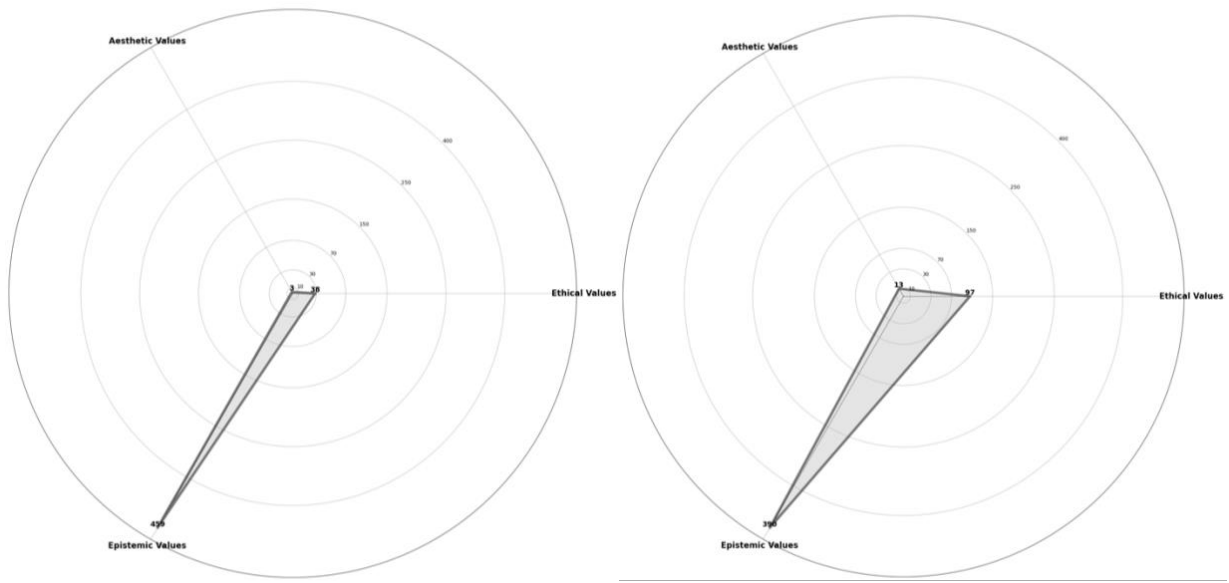

*Figure 10. Distribution of axiological perspectives. Left: City in computer science. Right: City in social science.*

## 5.6. Valorization dimension

Economic and social valorization are the dominant perspectives across all domains, with minimal focus on cultural valorization. In AI in computer science, economic valorization is the most dominant perspective with 253 instances, followed by social valorization with 231 instances. Cultural valorization is minimal with only 16 instances. AI in social science shows a similar pattern with social valorization leading at 306 instances, followed by economic valorization at 172, and cultural valorization at 22. In city in computer science, economic valorization is again dominant with 311 instances, followed by social valorization with 177 instances, and cultural valorization at 12 instances. City in social science shows a more balanced distribution with social valorization at 262 instances, economic valorization at 202, and cultural valorization at 36. These insights emphasize the need to integrate economic, social, and cultural benefits in AI and urban studies to address the diverse impacts of research and ensure that it contributes to the broader societal good.

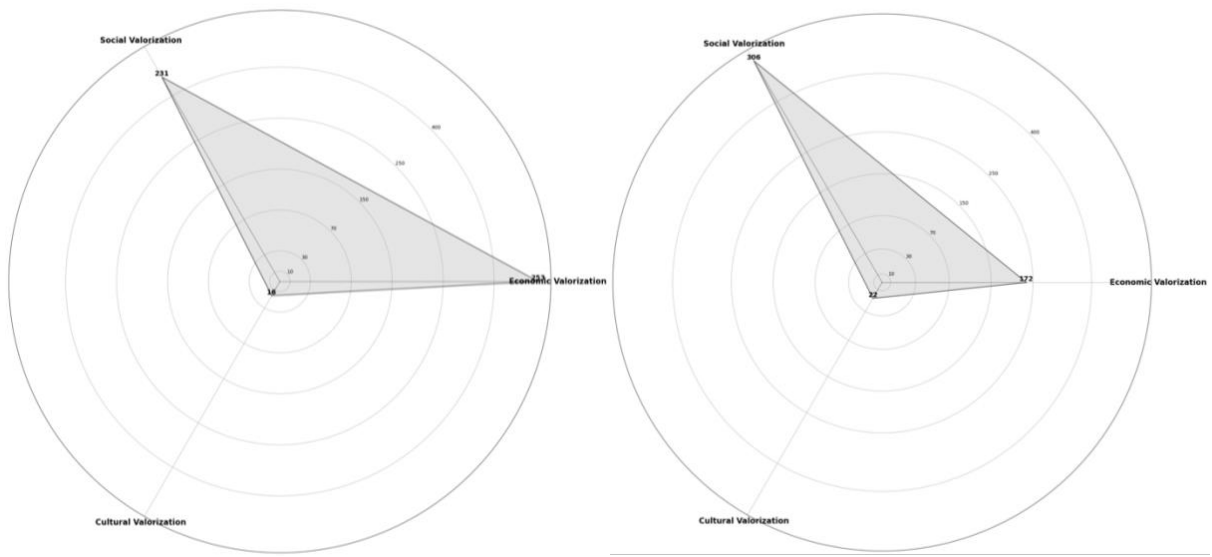

*Figure 11. Distribution of valorization perspectives. Left: AI in computer science. Right: AI in social science.*

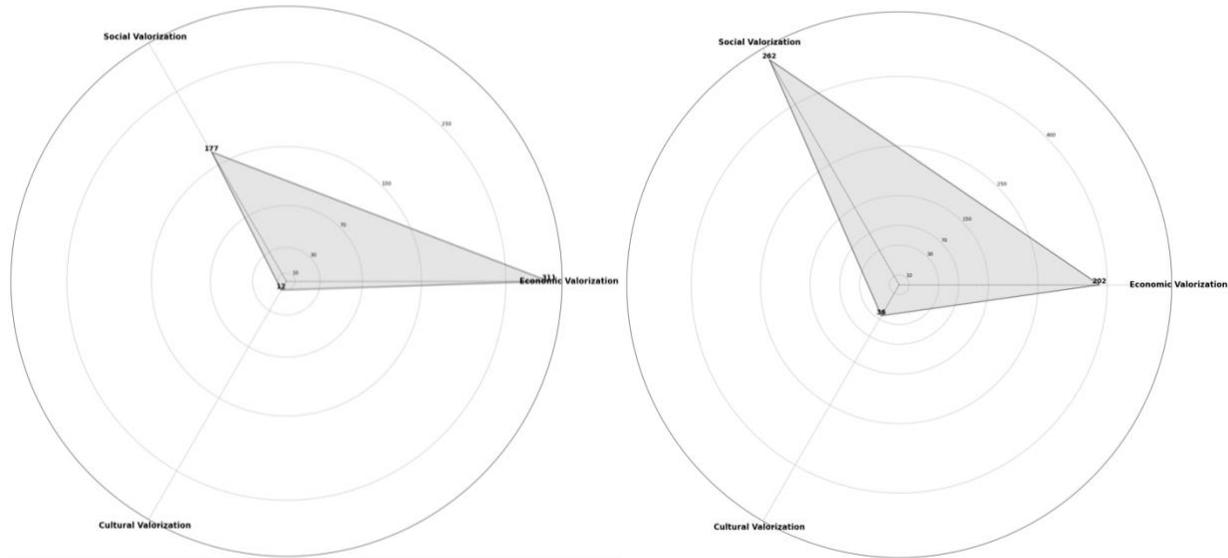

*Figure 12. Distribution of valorization perspectives. Left: City in computer science. Right: City in social science.*

## 6. Discussion and implications

The analysis identifies critical realism, positivism, analytical methods, and consequentialism as dominant perspectives within AI and urban studies, facilitating cross-domain validation. These viewpoints align with prevailing institutional norms in both computer science and social science, providing a common framework for interdisciplinary dialogue (Klein, 2017). Researchers aiming to produce broadly accepted intersectoral knowledge may find it advantageous to adopt these mainstream perspectives or engage with them critically to ensure their work resonates across disciplines.

In contrast, less prevalent perspectives—such as idealism and cultural valorization—offer the potential to incorporate innovative insights that address humanistic and contextual dimensions often neglected in AI-driven urban research (Costanza-Chock, 2020). The marginal representation of these viewpoints may result from systemic biases, entrenched disciplinary traditions, or limited publication outlets dedicated to transdisciplinary work (Vantard et al., 2023). However, integrating these underrepresented perspectives can enhance transdisciplinary research by emphasizing cultural, interpretive, and participatory aspects, thereby enriching the overall knowledge production process.

To foster effective transdisciplinary collaborations, several strategies emerge from reviewing relevant literature and practical experience. Reflexive alignment involves researchers critically assessing their inherent ontological and epistemological positions to understand how these perspectives facilitate or hinder engagement with other disciplines (Hadorn et al., 2008). This critical self-awareness enables researchers to identify and mitigate potential biases, fostering more effective interdisciplinary communication and collaboration. Stakeholder engagement is another essential strategy, whereby input from policymakers, practitioners, and community members is incorporated from the project's inception. This approach ensures that research questions and outcomes are relevant, ethical, and actionable (Fischer, 2000; Pratt, 2019). Additionally, iterative evaluation—implementing continuous feedback mechanisms and transparent evaluation criteria—can reduce biases against intersectoral projects, fostering recognition of diverse methodological and theoretical contributions (Klein, 2008; Wen et al., 2015). These strategies collectively support the development of transdisciplinary teams that enhance both the methodological rigor and social accountability of their research.

These strategies support the development of transdisciplinary teams that enhance both the methodological rigor and social accountability of their research. As AI becomes increasingly integral to policy and planning, it is necessary that collaborative efforts incorporate diverse epistemic and ontological perspectives to effectively address societal challenges, such as sustainable urban development and climate resilience (Hu et al., 2024).

## 7. Limitations

This study is subject to several limitations that may influence the generalizability and comprehensiveness of its findings. Firstly, the literature search was confined to English-language publications indexed in the Scopus database, potentially leading to the exclusion of relevant research published in other languages or indexed in different databases. This language restriction may result in an underrepresentation of diverse perspectives, particularly from non-English-speaking regions. Additionally, the selection of the 500 most cited papers per domain may inherently favor older publications with longer citation windows, possibly overlooking more recent but impactful studies. While a uniform citation cut-off was employed to mitigate disciplinary differences in citation practices, this approach may still bias the sample towards established theories and methodologies, thereby limiting the inclusion of emerging or innovative perspectives.

Moreover, the use of the GPT-4o Large Language Model for categorizing articles introduces potential biases and inaccuracies inherent in AI-driven classification processes. Automated categorization can be challenging due to the fluid and sometimes ambiguous definitions of perspectives across different fields, leading to inconsistent or erroneous classifications. The complexity of accurately mapping ontological, epistemological, methodological, teleological, axiological, and valorization dimensions further complicates the categorization process. Additionally, the exclusion of literature reviews, while intended to focus on primary research, may omit significant theoretical insights and conceptual frameworks that are pivotal for a comprehensive understanding of intersectoral research dynamics. Lastly, specialized subfields such as environmental psychology and human-computer interaction might be underrepresented if they were not adequately captured by the chosen keywords, potentially limiting the framework's applicability across all relevant areas of AI and urban studies. Despite these considerations, this approach is designed to ensure replicability and provide insight into prevailing research trends and perspectives in each domain.

## Conclusion

This study proposes and demonstrates a six-dimensional framework for evaluating intersectoral knowledge production at the intersection of AI, social sciences, and urban planning. By analyzing the 2,000 most cited papers (2014–2024) within these domains, we identified critical realism (ontological), positivism (epistemological), analytical methods (methodological), consequentialism (teleological), epistemic values (axiological), and social/economic valorization (valorization) as dominant perspectives. Researchers aiming for broader acceptance of intersectoral work can benefit from aligning, at least initially, with these prevalent stances. Nevertheless, less common perspectives—while presenting higher risks for acceptance—may offer crucial, context-specific insights and should not be dismissed outright.

From a transdisciplinary standpoint, the framework highlights the importance of reflexivity, methodological pluralism, and active stakeholder involvement. By bridging disciplinary divides and integrating non-academic actors, transdisciplinary research can produce innovative strategies to tackle the complexities of urban life, enhance ethical and cultural sensitivity in AI, and ultimately contribute to more inclusive societal outcomes. Future research should adapt and refine this framework for other contexts—such as health, sustainability, or global policymaking—and assess the long-term impacts of transdisciplinary projects on scientific advancement and real-world solutions.

*Competing interests:* The authors declare no competing interests.

*Data availability statement:* The datasets generated during the current study are available in a Hugging Face repository: https://huggingface.co/datasets/rsdmu/intersectoriality

*Ethical statements:* This article does not contain any studies with human participants performed by any of the authors.